# Comment on "Unraveling Photoinduced Spin Dynamics in the Topological Insulator Bi$_2$Se$_3$"

Y. D. Glinka

*Department of Physics and Astronomy, West Virginia University*
*Morgantown, West Virginia 26506-6315, USA*

Comment on a recent paper published in Physical Review Letters by M. C. Wang, S. Qiao, Z. Jiang, S. N. Luo, and J. Qi [Phys. Rev. Lett. **116**, 036601 (2016) - arXiv:1511.02994v2].

Although high-quality experimental results have been presented in a recent Letter [1], the conclusions made by Wang, Qiao, Jiang, Luo and Qi are not valid and provide the appropriate perspective to interpret ultrafast results in the broad family of two-dimensional materials. The authors have reported extremely short electron-phonon relaxation times of ~30 fs and ~300 fs, which characterize the bulk transverse/longitudinal optical (TO/LO-) phonon-mediated intraband cooling process for different phonon modes ($E_g^2$ and $A_{1g}^1$, respectively) in the topological insulator Bi$_2$Se$_3$. These short decay times, which have been reported only by the authors of Ref. [1] and by some of them in their previous work (~110 fs) [2], are even shorter than those known for the electron-optical-phonon relaxation in graphene (~200-500 fs) [3,4], despite the fact that the optical phonon energy in graphene (~196 meV) is at least 10 times larger than that in Bi$_2$Se$_3$ [5]. In contrast, the typical range of the electron-phonon relaxation times in single crystalline Bi$_2$Se$_3$ observed by other researchers is much longer (~2-4 ps for the $A_{1g}^1$ phonon mode [6-15] and 1.43 ps and ~1 ps for the $E_g^2$ and $A_{1g}^2$ phonon modes, respectively [10]). For Bi$_2$Se$_3$ thin films, the electron-phonon relaxation for the $A_{1g}^1$ phonon mode shortens to ~1.5 ps with decreasing film thickness in the range from 40 to 6 nm as a consequence of the more prominent effect of metallic-type Dirac surface states (SS) [11] and becomes even shorter (~700 fs) for thinner than 6 nm films [14] when direct intersurface coupling between opposite-surface Dirac SS [16] and quantum confinement effects [17] additionally contribute to the relaxation dynamics.

One exception is the short relaxation time in single crystalline $Bi_2Se_3$ of ~700 fs reported in Ref. [18], which has also been attributed to an optical phonon-mediated intraband cooling of conduction band electrons, despite oscillations associated with coherent $A_{1g}^1$ phonons were not observed in that study, in stark contrast to those observations in Refs. [6-8,10-14]. Consequently, this decay-time in Ref. [18] can result from other processes, for example, such as the nonequilibrium plasmon emission or Auger scattering in Dirac SS [19], which normally occur in graphene in the sub-picosecond range [20,21]. Nonetheless, even this short relaxation time is much longer than those attributed by the authors of Ref. [1] to the electron-phonon relaxation time in $Bi_2Se_3$. Moreover, they have suggested the existence of two electron-phonon relaxation processes occurring at electronic temperatures $T_e \gtrsim 600$ K (decay times 30 and 300 fs) and at $T_e < 600$ K (decay-time ~1 ps).

Because $Bi_2Se_3$ in the bulk is a polar semiconductor with a bandgap of $E_g$ ~0.3 eV [6-15], the electron-phonon relaxation in the conduction band should be considered in the frame of the polar Fröhlich interaction [22,23]. Using this concept, the electron-LO-phonon scattering time for the $A_{1g}^1$ phonon mode in $Bi_2Se_3$ has been estimated as ~31 fs [11]. Because the excess electron energy the authors used in their experiments is $\hbar\omega_{photon} - E_g$ ~1.25 eV (where $\hbar\omega_{photon}$ is the incident photon energy) [1] and the bulk LO/TO phonon energies are $A_{1g}^1$ (9.1 meV), $E_g^2$ (16.1 meV), and $A_{1g}^2$ (21.5 meV) [5], electrons should quickly cascade down to the conduction band minimum by emitting at least, for example, ~100 $A_{1g}^1$ phonons [11]. This cascading process immediately provides the electron-phonon relaxation time of ~3 ps, as reported in numerous publications [6-14]. Furthermore, $T_e = 600$ K (is equivalent to the bulk electron energy $1.5k_BT_e = 77.6$ meV, where $k_B$ is the Boltzmann constant) cannot be a unique temperature because it significantly exceeds LO/TO phonon energies in $Bi_2Se_3$ and therefore the emission of, for example, $A_{1g}^1$ phonons by hot electrons will continue until $T_e < 70$ K (9.1 meV). However, because phonon-assisted relaxation dynamics can be overlapped with the other carrier density dependent relaxation processes dominating at $T_e < 600$ K, such as carrier recombination [24], one can observe a crossover from one regime to the other, which occurs at $T_e = 600$ K [18].

The electron-phonon relaxation times in the bulk conduction band of $Bi_2Se_3$ on a time-scale of ~30 fs and ~300 fs seem hence inappropriate.

Instead, the relaxation trend discussed in Fig.1(a) of the Letter is similar to that reported previously for several semiconductor systems, such as GaAs [25] and Te [26], where the initial decay of the transient reflectivity signal has been associated with a macroscopic longitudinal polarization driven by the buildup of a photo-Dember field or by ultrafast screening of the depletion-layer field via the optical carrier "shock wave" injection process, which drives the coherent LO-phonon oscillations and therefore the initial decay cannot be treated separately from the entire oscillatory part, as the authors of the Letter nevertheless did and obtained unusually short electron-phonon relaxation times.